\documentclass[a4paper,12pt]{article}
\usepackage{epsfig}

\begin{document}

\def\beq{\begin{equation}}
\def\eeq{\end{equation}}
\def\bea{\begin{eqnarray}}
\def\eea{\end{eqnarray}}

\def\dofigure#1#2{\centerline{\epsfxsize=#1\epsfig{file=#2, width=7cm, 
height=4cm, angle=0}}}
\def\dofigureb#1#2{\centerline{\epsfxsize=#1\epsfig{file=#2, width=8cm, 
height=10cm, angle=-90}}}

\def\dofourfigs#1#2#3#4#5{\centerline{
\epsfxsize=#1\epsfig{file=#2, width=7.5cm,height=7.5cm, angle=0}
\hspace{0cm}
\hfil
\epsfxsize=#1\epsfig{file=#3,  width=7.5cm, height=7.5cm, angle=0}}

\vspace{0.5cm}
\centerline{
\epsfxsize=#1\epsfig{file=#4, width=7.5cm,height=7.5cm, angle=0}
\hspace{0cm}
\hfil
\epsfxsize=#1\epsfig{file=#5,  width=7.5cm, height=7.5cm, angle=0}}
}

\def\dotwofigsa#1#2#3{\centerline{
\epsfxsize=#1\epsfig{file=#2, width=8cm,height=3.2cm, angle=0}
\hspace{-1cm}
\hfil
\epsfxsize=#1\epsfig{file=#3,  width=8cm, height=7cm, angle=0}}
}

\def\dotwofigsb#1#2#3{\centerline{
\epsfxsize=#1\epsfig{file=#2, width=8cm,height=7cm, angle=0}
\hspace{0cm}
\hfil
\epsfxsize=#1\epsfig{file=#3,  width=8cm, height=7cm, angle=0}}
}

\def\dofig#1#2{\centerline{
\epsfxsize=#1\epsfig{file=#2, width=12cm,height=9cm, angle=0}
\hspace{0cm}
}}

\def\dofigc#1#2{\centerline{
\epsfxsize=#1\epsfig{file=#2, width=11cm,height=3.5cm, angle=0}
\hspace{0cm}
}}

\def\dofigb#1#2{\centerline{
\epsfxsize=#1\epsfig{file=#2, width=8cm,height=6.9cm, angle=0}
\hspace{0cm}
}}

\def\dofiga#1#2{\centerline{
\epsfxsize=#1\epsfig{file=#2, width=4cm,height=4cm, angle=0}
\hspace{0cm}
}}

\newcommand{\dedouble}{ \stackrel{ \leftrightarrow }{ \partial } }
\newcommand{\deR}{ \stackrel{ \rightarrow }{ \partial } }
\newcommand{\deL}{ \stackrel{ \leftarrow }{ \partial } }
\newcommand{\ci}{{\cal I}}
\newcommand{\ca}{{\cal A}}
\newcommand{\Wp}{W^{\prime}}
\newcommand{\vep}{\varepsilon}
\newcommand{\kk}{{\bf k}}
\newcommand{\pp}{{\bf p}}
\newcommand{\hs}{{\hat s}}
\newcommand{\proj}{\frac{1}{2}\;(\eta_{\mu\alpha}\eta_{\nu\beta}
+  \eta_{\mu\beta}\eta_{\nu\alpha} - \eta_{\mu\nu}\eta_{\alpha\beta})}
\newcommand{\projm}{\frac{1}{2}\;(\eta_{\mu\alpha}\eta_{\nu\beta}
+  \eta_{\mu\beta}\eta_{\nu\alpha}) 
- \frac{1}{3}\;\eta_{\mu\nu}\eta_{\alpha\beta}}

\def\lsim{\raise0.3ex\hbox{$\;<$\kern-0.75em\raise-1.1ex\hbox{$\sim\;$}}} 

\def\gsim{\raise0.3ex\hbox{$\;>$\kern-0.75em\raise-1.1ex\hbox{$\sim\;$}}}

\def\Frac#1#2{\frac{\displaystyle{#1}}{\displaystyle{#2}}}
\def\no{\nonumber\\}
\renewcommand{\thefootnote}{\fnsymbol{footnote}}

{\Large
\begin{center} 
{\bf 
Higgs boson plus photon production at the LHC: \\
a clean probe of  the b-quark parton densities
}
\end{center}}
\vspace{.3cm}

\begin{center}
Emidio Gabrielli$^{1,2}$,
$\;$ Barbara Mele$^{3}$, $\,$ and Johan Rathsman$^{1}$

$^1$\emph{High Energy Physics, Uppsala University, Box 535, 751 21 Uppsala,
Sweden }
\\
$^2$\emph{Helsinki Institute of Physics, P.O. Box 64 00014 University
of Helsinki, Finland}
$^3$\emph{INFN, Sezione di Roma,
and Dipartimento di Fisica, \\ Universit\`a di Roma `''La Sapienza", 
P.le A. Moro 2, I-00185 Rome, Italy}
\end{center}

\vspace{.3cm}
\hrule \vskip 0.3cm
\begin{center}
\small{\bf Abstract}\\[3mm]
\begin{minipage}[h]{14.0cm}
Higgs boson production in association with a high $p_T$ photon 
at the CERN Large Hadron Collider is analyzed,   in the framework of the MSSM model,
for the heavier neutral Higgs bosons.
The request of an additional photon in the exclusive Higgs boson final state selects  $b$-quark pairs among the  possible initial partonic 
states, since 
gluon-gluon  initial states are not allowed by C-parity conservation.
Hence, the measurement of  cross sections for  
neutral Higgs boson plus photon production can provide 
a clean probe of  the $b$-quark  density in the proton 
as well as of the $b$-quark Yukawa coupling.
The suppression of the production rates by the $b$-quark  electromagnetic coupling 
can be compensated by the enhanced Higgs boson Yukawa coupling to $b$'s 
in the large $\tan{\beta}$ regime. 
The  Higgs boson decay into  a  tau-lepton pair is considered, and  irreducible backgrounds with corresponding signal significances are evaluated.
\end{minipage}
\end{center}
\vskip 0.3cm \hrule \vskip 0.5cm
\vskip 0.3cm

The search for the Higgs boson, that is at
the origin of the electroweak symmetry breaking 
and fermion mass generation, remains one of the main tasks
of the Large Hadron Collider at CERN.
Although the Higgs boson
 is still eluding any direct experimental test,
electroweak precision tests requires,
for data consistency in the Standard Model (SM),
 a mass of the Higgs boson close to the LEP2 experimental bound of
$114$ GeV\cite{leptwobound}. 

A light Higgs-boson mass scenario is naturally
predicted in the framework of the
minimal supersymmetric extension of the SM (MSSM), 
with  a mass range close to the present experimental bound.
In the MSSM the Higgs boson sector contains two complex Higgs doublets, 
leading to two CP-even ($h,H$) and  one CP-odd ($A$) neutral Higgs bosons, and a pair 
of charged Higgs bosons. At tree level, the Higgs sector of MSSM is 
fully specified by 2 parameters (e.g.,  $\tan{\beta}$, the ratio of the
up- and down-Higgs doublet vacuum expectation values, and the mass
of the pseudoscalar $m_A$), while higher-order radiative corrections 
to the effective potential, and hence to the masses, 
depend on other relevant SUSY parameters.
Large $\tan{\beta}$ scenarios are favored in the  MSSM,
since they can easily provide both a viable Dark Matter candidate 
and large radiative corrections to the lightest Higgs boson mass.
Other indications in this direction comes
from the SUSY contribution to the anomalous magnetic moment of the muon 
$(g-2)_{\mu}$\cite{H-MSSM}.

The $A/H$ boson Yukawa couplings to  $b$-quarks and tau leptons are 
also  enhanced by a large $\tan{\beta}$.
The CDF and D0 collaborations at Tevatron have
recently  analyzed the process $p\bar{p}\to \phi \to \tau^+\tau^-$, 
where $\phi=\{h,H,A\}$ \cite{cdf,d0}.
Although  no evidence for a Higgs boson signal
has been reported, the  limits obtained for  $\tan{\beta}$ and
$m_A$ by the CDF collaboration turn out  to be weaker 
than expected,
 in the mass region  $m_A\sim120-160$ GeV. 
This could be a hint
of a  MSSM scenario with $m_A\sim 160$ GeV and $\tan{\beta}\gsim 45$ \cite{0706.0977}.

In the large $\tan{\beta}$ regime,
the $b$-quark fusion process becomes one of the main mechanisms for heavy Higgs boson production at hadron colliders. While, in the SM,  the  $b\bar b\to h$
cross section at the LHC is at least two orders of magnitude smaller that the main gluon fusion  cross section, in the MSSM the $b$ fusion  rates are comparable to the $gg\to A/H$ ones, even at moderate values of
$\tan{\beta}$  \cite{hep-ph/0607308}.
In particular, for $m_{A,H} \gsim 150$ GeV and $\tan{\beta} \gsim 5$, the bulk  of the 
$A/H$ bosons
produced at the LHC comes either from $gg \to A/H$ or from $b\bar b\to A/H$,  the latter contributing dramatically to the Higgs discovery especially at large 
values of $m_{A,H}$ and $\tan{\beta}$.

The $b$ fusion process is particularly sensitive not only to the Higgs coupling to the $b$ quark, but also to the actual composition of the proton in terms of the 
$b$-quark partons. 
At present, the $b$-quark parton density is derived perturbatively from the gluon parton density, and there is no direct measurement of it. In fact, the uncertainty in  the gluon content of the proton propagates to the $b$ component
prediction.
In the SM, one is planning to measure the 
$b$-quark parton densities through processes containing one $b$-jet in the final state, such as $bg\to b Z/b\gamma$ \cite{hep-ph/0601164}. 
The $bg\to b Z/b\gamma$ cross sections are  linearly dependent on the $b$-quark densities. A measurement of the  $b\bar b\to h$ production rate would be remarkably more sensitive, but there is no hope  in the SM of separating the $b\bar b\to h$ signal from the 
$gg\to h$ production.

On the contrary,  in the MSSM, the inclusive heavy Higgs production is predicted to be quite sensitive to  the  $b\bar b\to h$ 
contribution, although  a non-negligible contamination from  $gg\to h$
could make the extraction of the information on the $b$-quark densities more involved for large portions of the MSSM parameter space.

The aim of the present work is to 
study the exclusive process 
$pp\to \phi \,\gamma$ ,
 with
$\phi=\{h,H,A\}$, at the LHC, where a high $p_T$ photon is required in the final state.
The advantage to be gained from the photon is that at leading order
the exclusive $\phi\;\gamma$ final state selects the $b\bar{b}$ 
 initial state through the channel $b\bar b \to \phi\,\gamma$ 
shown in Figure~1.
Indeed, the $gg\to \phi\,\gamma$ transition 
is forbidden at any order in perturbation theory by 
C-parity conservation.
One can  easily check, at one loop, that
 the box amplitude for $gg\to \phi \,\gamma$ exactly vanishes, even 
for off-shell external states.
Hence, possible next-to-leading corrections coming from 
extra soft gluons emitted from the initial gluons vanish, too,
while gluon emission from the box vanishes in the soft limit due to
gauge invariance.
Hence, the vanishing of the exclusive $\phi \,\gamma$ production
from gluon fusion is stable under radiative corrections.
Note that the process $gg\to \phi \,Z$ does not share the same property \cite{ggzgamma}.
\begin{figure}[tpb]
\dofigc{3.1in}{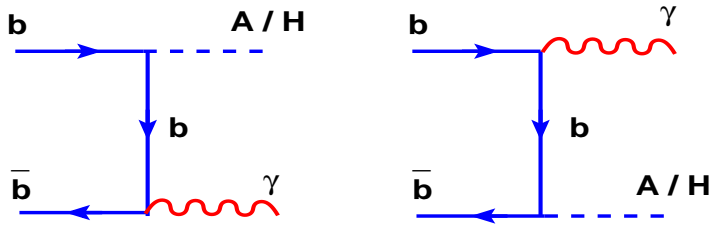}
\vspace{-.3cm}
\caption{\small Feynman diagrams for the process $b\bar{b}\to A/H\, \gamma$.
}
\label{Fig1}
\end{figure}
Note also that when studying the $A/H\to \tau\tau$ signal with an additional
$b$-tagging, there is in principle a large contamination arising from the
$gb$ initial state. In addition, in this case, tagging a
 $b$-jet gives rise to double counting in the Monte Carlo simulation of
events \cite{johan}.

Of course, the presence of an additional photon in 
$b\bar b\to \phi \,\gamma$ 
 suppresses the corresponding Higgs production rates, due to both the electromagnetic coupling $\alpha_{em}$ and the $b$-quark electric charge $Q_b=-1/3$. 
 Nevertheless, we will see that  the large-$\tan{\beta}$ enhancement in the $A/H$ production can 
partly compensate this suppression, and leads to measurable rates.

In large-$\tan{\beta}$ scenarios,
the most promising signature for the process
$pp\to \phi\,\gamma$ is the one arising from 
the decays $A/H\to \tau\tau$. Indeed, one has a branching ratio 
$BR(A/H\to \tau\tau)\simeq 10\%$, and, on the other hand,
the irreducible background for the  $\tau\tau\,\gamma$ final state
has a purely electroweak origin, and will be shown to be well under control.

The $\tau\tau$ signature in Higgs boson production at the LHC
has been extensively studied both in the SM Higgs boson production via vector boson fusion and in the inclusive MSSM  A and H boson 
production \cite{hep-ph/0203056}.
It turns out to be in general extremely promising, providing a discovery channel in the SM case for quite light  Higgs-boson masses \cite{hep-ph/0402254}. 
Also, a few percent accuracy on the measurement of the $A/H$ masses 
in the MSSM is expected on the basis of the $\tau\tau$ invariant mass reconstruction
\cite{0704.0619}.

Although the corresponding Higgs boson  $BR$ in $\tau\tau$ 
is in general one order of magnitude smaller than that of 
the dominant decay mode into $b\bar{b}$, the possibility
of tagging $\tau$ leptons allows a drastic reduction of the background. On the other hand, despite the fact that 
undetected neutrinos are present among the decay products of the 
$\tau$'s, one can  reconstruct
the complete $\tau\tau$ invariant mass, whenever  the two $\tau$ lepton are neither 
back-to-back nor collinear in the laboratory frame~\cite{KEllisNPB297-221(1988)}. 
Requiring a large $p_T$ photon in the final state of the process 
$pp\to \phi\,\gamma$ naturally produces the required kinematical $\tau\tau$ configuration.

Note that  the $A$ and $H$ bosons masses,
as well as their couplings to fermions and decay widths,
tend to be degenerate at large $\tan{\beta}$ and large $m_A$.
As a consequence, the experimental separation of the two 
signals arising from $A$ and $H$ will be challenging. In the following analysis,
we will then combine the statistical samples corresponding to the 
$A$ and $H$ resonances.

The analytical expression for the 
tree-level  differential cross section $\sigma_{\phi}\equiv 
\sigma(b\bar{b}\to \phi \,\gamma)$, for $\phi=\{h,H,A\}$, is given by
\bea
\frac{d\hat{\sigma}_{\phi\gamma}}{d\hat{t}}(\hat{s},\hat{t})=
\frac{\alpha_{em}\, Q_b^2\, \lambda_{\phi}^2}{4 N_c\,(1-4r_b)}\left\{
\frac{F_1^{\,\phi}(\hat{s})}{(\hat{t}-m_b^2)(\hat{u}-m_b^2)}\right.
+\left. F_2^{\,\phi}(\hat{s})\left(\frac{1}{(\hat{t}-m_b^2)^2}
+\frac{1}{(\hat{u}-m_b^2)^2}\right)\right\},
\label{xsection}
\eea
where
\bea
F_1^{\,\phi}(\hat{s})=(1-r_{\phi})^2+
2\,(r_{\phi}-r_b\,\xi_{\phi})(1-2r_b)\, , \;\;\;\; 
F_2^{\,\phi}(\hat{s})=-2\,r_b\,(r_{\phi}-r_b\,\xi_{\phi})\, ,
\label{F12}
\eea
$N_c$  is the color number, 
$r_{i}=m_{i}^2/\hat{s}\;$ ($i=\phi,b$), 
 ~$\xi_{\,h,H}=4$, and $\xi_A=0$.
The Yukawa couplings $\lambda_{\phi}$ are given by 
$\lambda_{h}=\lambda^{SM} (-\sin{\alpha}/\cos{\beta})$,
$\lambda_{H}=\lambda^{SM} (\cos{\alpha}/\cos{\beta})$, and 
$\lambda_{A}=\lambda^{SM} \tan{\beta}$, 
being $\lambda^{SM}=m_b/v$, with 
$v$ the electroweak vacuum expectation value.
 The
 Mandelstam variables are defined as 
$\hat{t}=(p_b-p_{\gamma})^2$,
$\hat{u}=(p_{\bar{b}}-p_{\gamma})^2$, $\hat{s}=(p_b+p_{\bar{b}})^2$.
Finally, 
since (at tree-level) 
$\sin{2\alpha}=-\sin{2\beta}\, (m_H^2+m_h^2)/(m_H^2-m_h^2)$,
at large $\tan{\beta}$   
the cross sections for the $H$ and $A$ boson  production
are  much larger than the  $h$ boson cross section.

In the $m_b\to 0$ limit,  in Eq.(\ref{xsection}) 
one has  $F^{\,\phi}_1(\hat{s})\to(1+r_{\phi}^2)$ ,
$F^{\,\phi}_2(\hat{s})\to 0$,  and the scalar and pseudoscalar boson cross 
sections have the same expression in terms of  Yukawa couplings
and masses, due to the chiral symmetry restoration.

We also take into account the leading-order 
QCD corrections to the Yukawa couplings, by using the
running $b$-quark mass $m_b(\mu)$, in the Yukawa coupling,
evaluated at a scale 
$\mu\sim m_{\phi}$. 
For $m_b^{\rm pole}=4.6$ GeV, one has 
$m_b(\mu)\simeq 2.81$ and 2.56 GeV, for  
$\mu\simeq 150$ and 500 GeV, respectively.
These results have been obtained using the HDECAY program~\cite{Hdecay},
also used to calculate the masses and widths of the Higgs
bosons.

In order to obtain the inclusive cross sections for the process $pp\to \phi\,\gamma$, we impose a minimal cut 
$p_T^{\rm cut}\gsim 30$~GeV on the $\gamma$ 
transverse momentum. The latter also regularizes soft and 
collinear singularities arising in the $m_b\to 0$ limit.

The $pp$ cross section 
$\sigma_{\phi\, \gamma}$, after integrating over
$p_T^{\gamma} > p_T^{\rm cut}$, is then 
\beq
\hspace{-0.4cm}
\sigma_{\phi\, \gamma}
=
\int_{x_1^{\rm c}}^1  \! dx_1 \!  \int_{x_1^{\rm c}/x_1}^1 \! \!  \!  \! \! dx_2 
\, \left[f_b(x_1) \, f_{\bar b}(x_2) +c.c. \right] 
 \, \hat{\sigma}_{\phi\gamma}(\hat{s})  \, ,
\label{xsectionpp}
\eeq
where $\hat{s}=x_1x_2\, S$, 
$x_1^{\rm c}\equiv
\left(p_T^{\rm cut}+\sqrt{(p_T^{\rm cut})^2+m_{\phi}^2}\right)^2/S$,  
$\sqrt{S}$ is the $pp$ center-of-mass energy,
and
$f_b(x)$ is the $b$-quark parton density in the proton. In 
Eq.~(\ref{xsectionpp}), 
the inclusive partonic cross section $\hat{\sigma}_{\phi\gamma}(\hat{s})$
can be obtained by
 integrating Eq.~(\ref{xsection})
\beq
\hat{\sigma}_{\phi\gamma}(\hat{s})=\frac{\alpha_{em}\, 
Q_b^2\, \lambda_{\phi}^2} {2\, N_c\,\hat{s}}\left\{
\, \frac{1+r_{\phi}^2}{1-r_{\phi}}\, 
 \log{\left(
\frac{1+\sqrt{\Delta}}{1-\sqrt{\Delta}}\right)}\right.
-\left.
2\,\frac{m_b^2}{(p_T^{\rm cut})^2}\, \sqrt{\Delta}\; 
r_{\phi}\left(1-r_{\phi}\right) \right\} \,,
\label{total}
\eeq
where
$
\Delta=1-4\,(p_T^{\rm cut})^2 /
[\hat{s}(1-r_{\phi})^2]\, .
$
In  Eq.(\ref{total}),
we retained the leading contributions in $m_b$, 
that are of order ${\cal O}[m_b^2/(p_T^{\rm cut})^2]$.
They arise by integrating the 
terms $m_b^2/(t-m_b^2)^2$ and $m_b^2/(u-m_b^2)^2$ in the 
differential cross section in Eq.(\ref{xsection}), and have an impact of the order of a few permil for 
$p_T^{\rm cut}\simeq30$ GeV.
Note that, in the inclusive cross section, 
 $m_b^2/(t-m_b^2)^2$ and $m_b^2/(u-m_b^2)^2$ terms give also finite contributions that 
survive the $m_b\to 0$ limit after integration. 
 This mass discontinuity is due to a 
chirality-flip mechanism induced by collinear 
photons~\cite{chirality-flip}.

\begin{figure}[tpb]
\dofigb{3.1in}{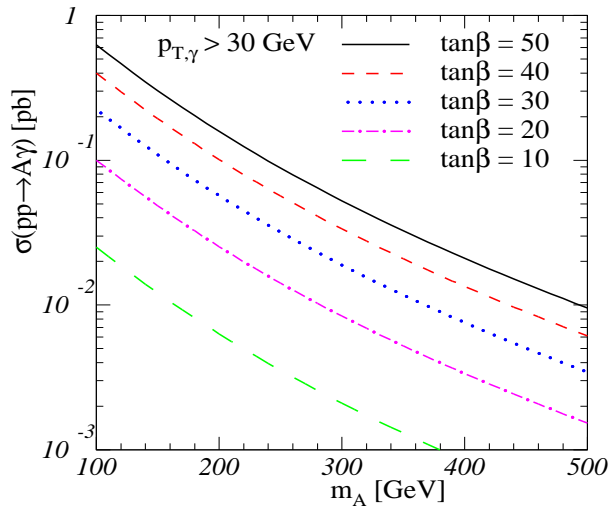}
\vspace{-.3cm}
\caption{\small Total cross section (in pb) for the process
$pp\to A\gamma$ at the LHC, for $p_T^{\gamma}> 30 $ GeV, and  several values of 
$\tan{\beta} =10,20,30,40,50$ (from lower to higher curve).
}
\label{Fig2}
\end{figure}

The inclusive cross section  for
$pp\to A\,\gamma$ versus $m_A$ and  $\tan{\beta}$  is shown 
 in  Figure~\ref{Fig2}, for $p_T^{\rm cut}\simeq30$~GeV. 
We use the parton density set CTEQ6L1~\cite{CTEQ}, with the factorization scale  $\mu_F\simeq m_A/2$.
Cross sections of a few hundreds  fb's are obtained for a relatively light
$m_A$ and large $\tan{\beta}$ values. 
Assuming $p_T^{\rm cut}=30, 40, 50, 60$ GeV,  for 
$m_A= 150\, (500)$~GeV and 
$\tan{\beta}=40$, one gets 
$\sigma = 194 , 116 , 73.6 , 49.0 \; (6.12 ,4.36 , 3.26 , 2.50)$~fb,
respectively.
 Similar results hold for the heavy scalar Higgs boson 
 process $pp\to H\,\gamma$, for
 $m_H\simeq m_A$ and same $\tan{\beta}$.
 
 Note that the corresponding $p\bar p\to A~\gamma$
 cross section at  Tevatron 
 turns out to be  2.5~ fb,  for $p_T^{\rm cut}\gsim 20$~GeV,
 if $m_A= 150$ GeV and $\tan{\beta}\simeq 50$.
 
We now present the results of a first analysis of the $\tau\tau$ signature corresponding to the process 
$pp\to \phi\, \gamma \to \tau^+\tau^-\gamma$ in the MSSM
at the LHC. It aims to separate the $b\bar b\to \phi\, \gamma$ signal from the main irreducible background corresponding to the $\tau$ pair production via
an off-shell $Z/\gamma$ associated to a high $p_T$ photon.
The spurious signal coming from the channels $gg,b\bar b\to \phi\to \tau^\ast\tau\to \tau \tau \gamma$, mediated by an off-shell $\tau$
that radiates a photon,
has also to be kept under control, since only the resonant 
$\phi\to\tau\tau$ production associated to a high $p_T$ photon can disentangle the
initial $b\bar b$ partonic state.

A much more detailed presentation than can be fit into this paper will be given in \cite{the-three-of-us?}.

The $pp\to \tau\tau \gamma$ cross sections for signal and SM backgrounds  have been calculated 
 by using the full tree-level matrix element through CompHep 
 \cite{comphep}, 
and also 
cross checked via our analytical results.
Apart from 
the $\tau$ pair production via
an off-shell $Z/\gamma$ associated to a high $p_T$ photon,
a further irreducible SM backgrounds is given by 
the channel $pp\to W^+W^- \gamma$,
when  $W^{\pm}\to\tau^{\pm}\nu $.
We checked that the latter contribution to the $\tau \tau 
\gamma$ signature is negligible, even including the other
leptonic $W$ decays that can fake taus decays. 
Indeed, we find $\sigma(pp\to W^+W^- \gamma)\times
Br(W\to l\nu_l)^2\simeq 14$ fb, for $p_T^{\gamma}\gsim 30$ GeV, which will be further reduced by  cuts on the $\tau\tau$ invariant
mass, etc. .

Due to the mass and coupling degeneracy discussed above,
the signal corresponding to $b\bar b \to A\,\gamma\to \tau\tau\gamma$
and $b\bar b \to H\,\gamma\to \tau\tau\gamma$
will be added coherently,  increasing the total production rate
by  a factor two.

In counting the expected number of events for a given integrated luminosity, 
we assume quite conservatively a detection efficiency  for the tau pair
of  $\epsilon_{\tau\tau}\simeq 0.2$. This takes into account
both the leptonic $\tau\to \ell \,\nu_{\tau}\nu_{\ell}$  
(35\%) and the one-prong 
hadronic $\tau\to h \nu$ (50\%) signatures  for each of the two taus,
assuming  ID efficiency  $\sim90\%$ and $\sim25\%$, 
respectively (giving a detection efficiency $\epsilon_{\tau}\simeq 0.44$ on a single tau)~\cite{Rainw}.
Note that the doubly hadronic decays contributes only by about 0.016
to $\epsilon_{\tau\tau}\simeq 0.2$ in this case.

In order to optimize the suppression of any  $\tau\tau\gamma$ final state not coming from the resonant 
$b\bar b \to \phi\,\gamma\to \tau\tau\gamma$,
we impose the following kinematical cuts :
\\
${\bf -}$ $p_T^{\gamma}\, > \, 30, 30,40,50$~GeV, for $m_A=150,200,300,500$ GeV,  respectively;
\\
${\bf -}$
$0.9\; m_A < m_{\tau\tau} < 1.1\; m_A~~$ on the $\tau\tau$ invariant mass ;
\\
${\bf -}$
$p_T^{\tau^{\pm}} >20 {\rm GeV},~
|\eta_{\gamma}| < 2.5,~
|\eta_{\tau^{\pm}}| < 2.5 $, where $|\eta_i|$ is the $i$-particle pseudo-rapidity;
\\
${\bf -}$
$\Delta R_{\gamma \tau^{\pm}} > 0.7,~~ 
\Delta R_{\tau \tau} > 0.7,~~ \Delta\, \phi_{\tau \tau} < 2.9$,  where
$\Delta R_{ij}=\sqrt{(\Delta \eta_{ij})^2+ (\Delta \phi_{ij})^2}$,
and $\Delta \eta_{ij}$ and $\Delta \phi_{ij}$ stand for 
the difference in pseudo-rapidity and azimuthal angle of 
the particle $i$ and $j$, respectively.

\begin{figure}[tpb]
\dotwofigsb{3.1in}{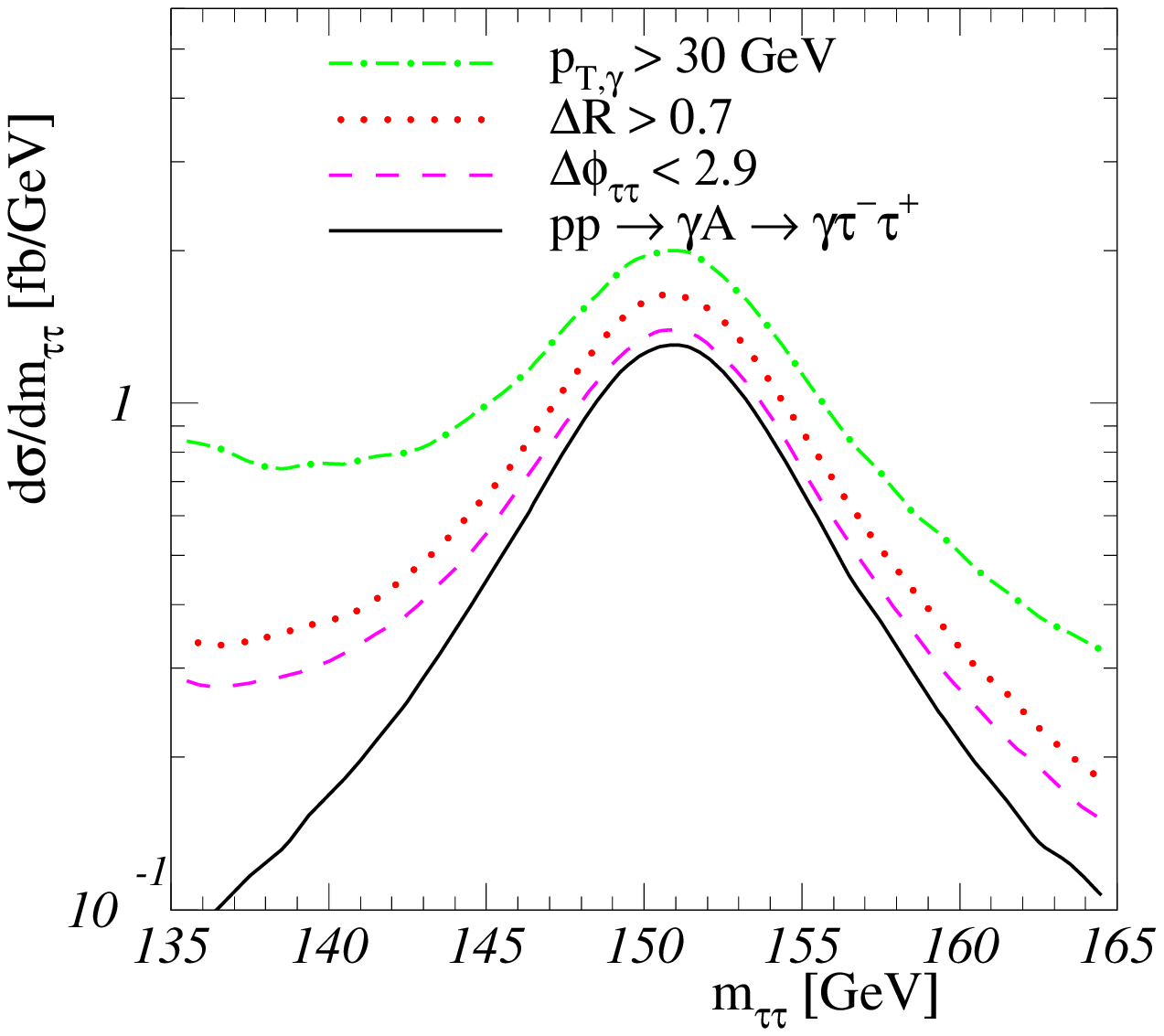}{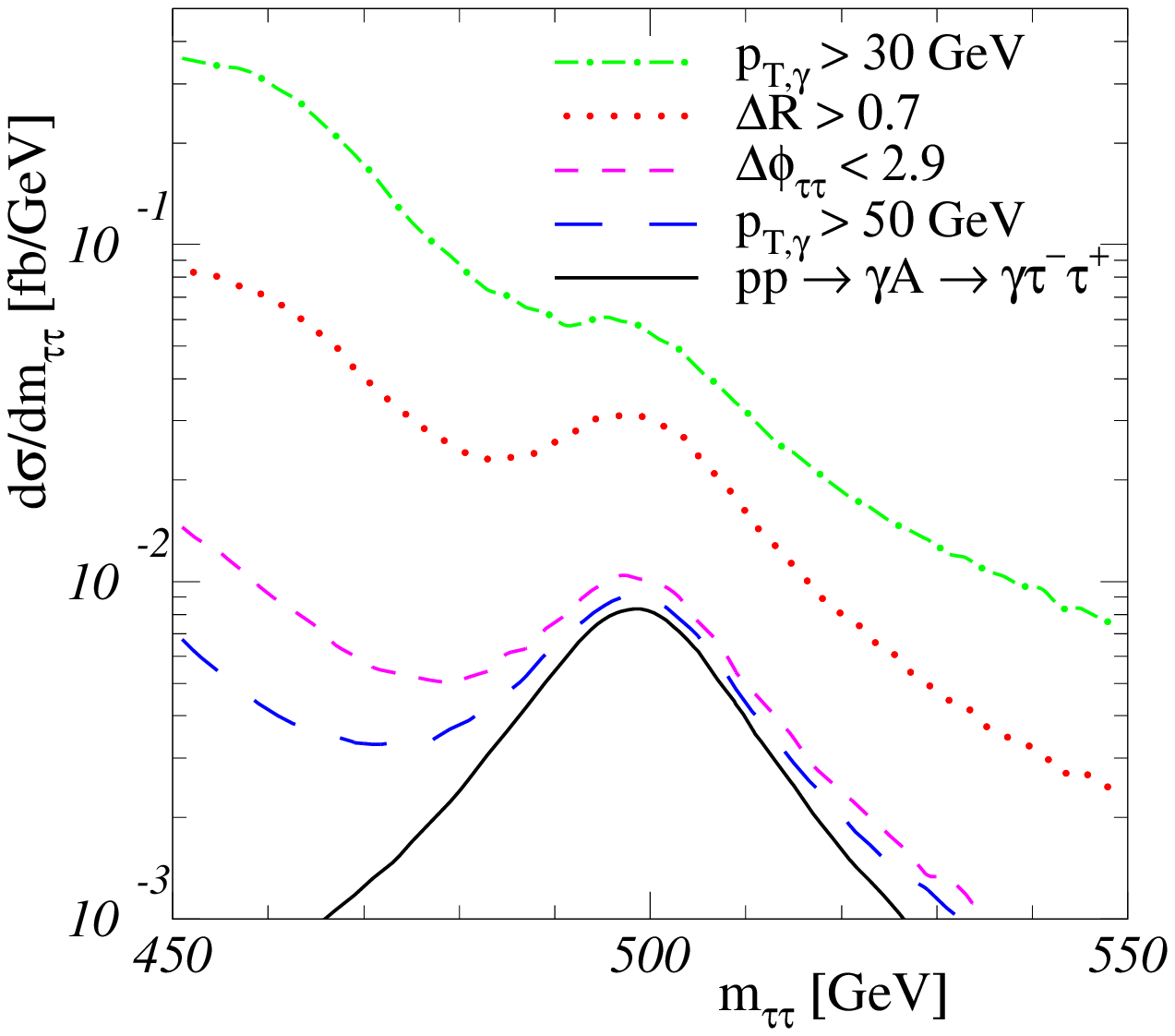}
\vspace{-.3cm}
\caption{\small Differential cross sections $d\sigma/dm_{\tau\tau}$ 
at the LHC for the 
process $b\bar b\to  \tau \tau \gamma$, at $\tan\beta=50$ and 
$m_A=150$ GeV (left),  500 GeV (right), 
after sequential application of the
cuts indicated (with $\Delta R\equiv\Delta R_{ij}$).  
Cuts on pseudo-rapidities and transverse momenta are 
detailed in the text.}
\label{Fig3}
\end{figure}

The $\Delta R_{ij}$ cuts guarantee the needed spatial isolation 
  to  identify particles experimentally.
  On the other hand, both the $\Delta R_{\gamma \tau^{\pm}}>0.7$ and 
  the $\Delta\, \phi_{\tau \tau} < 2.9$ cuts suppress the kinematical configurations where the photon is almost collinear to one of the $\tau$.
  The latter are typical of photons emitted off taus, like in $gg,b\bar b\to \phi\to \tau^\ast\tau\to \tau \tau \gamma$, that we want to deplete.
  Adding the cut $\Delta\, \phi_{\tau \tau} < 2.9$ after imposing
  $\Delta R_{\gamma \tau^{\pm}}>0.7$ has a modest impact at $m_A\sim 150$~GeV,
  but is quite effective at larger masses (see  
  Figure~\ref{Fig3}).
  Imposing the $\tau\tau$ acollinearity condition 
  $\Delta\, \phi_{\tau \tau} < 2.9$ also guarantees  the reconstruction of the 
  complete invariant mass for the   $\tau\tau$ system despite the presence of invisible neutrinos in the $\tau$ decay products 
  \cite{Rainwater-Zeppenfeld}.
  
  Figure~\ref{Fig3} shows, for the $b$-fusion process $b\bar b\to  \tau \tau \gamma$ mediated by $A$ and $H$,
the effect of the cuts on $\Delta R_{\tau \gamma}$, 
$\Delta R_{\tau \tau}$, and $\Delta \phi_{\tau \tau}$ in 
dramatically reducing (especially at large $m_{A,H}$) 
the contribution of  
$\gamma$ emission off taus, that shifts $m_{\tau\tau}$ from the $m_{A,H}$ resonance.
After applying all cuts, one gets a substantial 
 purity of the $b\bar b\to \phi\,\gamma$ signal.  
\begin{table}[th]
\begin{center}
\begin{tabular}{||c||c|c|c|c|c|c|c|c||}
\hline 
$\tan{\beta}$  & \multicolumn{2}{c|}{20}   & \multicolumn{2}{c|}{30}  & \multicolumn{2}{c|}{40}  &  \multicolumn{2}{c||}{50}  \\
\hline \hline
$m_A$  & $\sigma_S$ (fb) & ${\cal S}$ &  $\sigma_S$ (fb) & ${\cal S}$  &  $\sigma_S$ (fb)&  ${\cal S}$&
$\sigma_S$ (fb)&  ${\cal S}$
 \\ \hline
150 & 5.58 & 7.3 & 12.5 & 13 & 22.1 & 19 & 34.5 & 24  \\ \hline
200 & 3.00 & 5.3 & 6.81 & 9.5 & 12.3 & 14 & 19.9 & 18 \\ \hline
300 & 0.727 & 2.4 & 1.67 & 4.5 & 3.08 & 6.7 & 5.03 & 9.1 \\ \hline
500 & 0.0981 &0.72 & 0.238 & 1.5 & 0.456 & 2.4 & 0.768 & 3.4 \\ \hline
\end{tabular}            
\caption{\label{table1} \small Signal cross section $\sigma_S$ 
for the $b$-fusion contribution to the  
process $pp\to \tau^+\tau^-\,\gamma$ at the LHC,
and corresponding
significance ${\cal S}=n(S)/\sqrt{n(S)+n(B)}$ 
versus $m_A$ (in GeV) and $\tan{\beta}$ 
($n(S)$ and $n(B)$ stand for the number of
events for signal and irreducible background, respectively).
An integrated luminosity of ${\cal L}=100~ {\rm fb}^{-1}$ and
a tau-pair reconstruction efficiency $\epsilon_{\tau\tau}\simeq0.2 $ are 
assumed in the ${\cal S}$ evaluation.
Kinematical cuts are defined in the text.
}
\end{center}
\end{table}
In particular,  the $\gamma$ radiation off $b$'s gives $83\%$ 
($66\%$) of the observed rate,
 for $m_A=150$ (500) GeV, at $\tan\beta=50$.

The  $b$-fusion cross sections $\sigma_S$ 
(mediated by both $A$ and $H$ bosons), 
 contributing to the 
$pp\to \tau\tau\, \gamma$ rate after the above cuts, 
are reported in 
Table~1.
For the same choice of cuts,
 the $Z^*/\gamma^*$ 
background cross sections  are  
$\sigma_B=6.10,~ 3.44,~ 1.12,~ 0.270$~ fb, 
for $m_A=150,~ 200,~ 300,~ 500$ GeV, respectively. Values for 
the significance  ${\cal S}=n(S)/\sqrt{n(S)+n(B)}\gsim 5$ 
(where $n(S)$ and $n(B)$ stand for the number of
events for signal and irreducible background, respectively)
are obtained  for $m_A\lsim 300$  GeV and $\tan{\beta}\gsim 30$, 
with an integrated luminosity of ${\cal L}=100~ {\rm fb}^{-1}$,
and the tau-pair detection efficiency $\epsilon_{\tau\tau}\simeq0.2 $.

In order to get a feeling of the sensitivity of the signal rates
to the present estimate of the error in the $b$-quark parton density,
we  computed the signal rates for several PDF sets among the
ones present in LHAPDF \cite{lhapdf}, and found
a variation of at most about 20\%. However, one should keep in mind
that the actual uncertainty on the PDF could be quite larger than
the one given by the present LHAPDF set \cite{Thorne:2007fe}.

Another source of theoretical uncertainty in our calculation comes from
the SUSY corrections to the b-Yukawa coupling that we
are not taking into account, and which are proportional to $1/(1+\Delta_b)^2$.
The term $\Delta_b$ depends also on further SUSY parameters (see 
\cite{carena} for more details).
However, although these corrections turn out to be in general quite large, 
they cancel out when the $\phi \to \tau\tau$ decay is considered, 
leaving only a residual dependence $\approx 1/[(1+\Delta_b)^2+9]$, just as 
is the case for the corresponding process without a photon~\cite{carena}.

Clearly, how well one will be able to determine both the
$b$-quark parton density
and the overall coupling $(\tan{\beta})^2$ will crucially depend on the
actual experimental precision on the measurement of the 
$b\bar b \to \phi\,\gamma$ cross section. We believe that both a more 
sharp theoretical prediction, i.e. the inclusion of 
higher order QCD corrections,
and a full experimental simulation of the signatures corresponding
to the signal and background, will be needed to assess the actual potential
of this process. 

Presently, studies have been carried out
in order to estimate the uncertainty on the measurement of $\tan{\beta}$
derived from the $b\bar{b}\phi$ production (see \cite{PTDR}). 
This uncertainty is expected to be quite large at the moment.
Assuming a comparable uncertainty for the value of 
the $\tan{\beta}$ coupling in the process $b\bar b \to \phi\,\gamma$ would
prevent the extraction of the $b$-quark density in a precise way.
However, one could think of new strategies aimed to optimize the
measurements of the relevant quantities coming from different processes.
For example, one could consider the ratio of the 
$b\bar b \to \phi\,\gamma$  cross section over
the $b \bar{b} \phi$ production cross section where
the dominant $(\tan{\beta})^2$ dependence (and uncertainty) would automatically 
drop off. One could then obtain an observable with higher sensitivity to 
the $b$-quark parton density.

In conclusion, we believe that the process $b\bar b \to \phi\,\gamma$,
possibly combined with other observables, could help in the determination
of both the $b$-quark parton density and the $b$-quark Yukawa coupling.
In order to establish its actual potential, further 
theoretical and experimental studies will be needed.

\

\ 

We thank N.~Mahmoudi, G.~Ingelman and O.~St{\aa }l for useful discussions.
E.G. and J.R. acknowledge financial support from the G\"oran Gustafsson Foundation.
B.M.'s  research was partially supported by the RTN European Programmes 
MRTN-CT-2006-035505 (HEPTOOLS, Tools and Precision Calculations for Physics 
Discoveries at Colliders), and  MRTN-CT-2004-503369 (Quest for Unification).

\end{document}